# VOLUMETRIC (CUBIC) OCTONION SIGMA- MATRICES AND THEIR PROPERTIES

Sergei Yakimenko

**Abstract.** This work rests upon the certainty that only fields of real and complex numbers, quaternions and octonions have algebras of all four arithmetical operations. Also quaternions are good to represent 3-dimensional Euclid space and 4-dimensional Minkowski space. Moreover algebra of Pauli classical sigma-matrices isomorphic to a split quaternion algebra.

In this article author supposes that the hyperspace can be described with octonions and if 4-dimensional Minkowski space is a projection of the hyperspace, consequently equations for 4-dimensional space must be projections of hyperspace equations. Therefore he obtains a new object as volumetric analog of Pauli sigma-matrices for 8 (or maybe 24)-dimensional hyperspace. Also the author represents general form of a volumetric Dirac equation solution.

As is well known nonassociativity is the main distinctive feature of octonions. But square matrices cannot be nonassociative [3]. Therefore the author had to introduce cubic sigma-matrices. From another hand a volumetric object has 3 indices, as a result of it one cannot introduce multiplicative operator for cubic matrices, their projections only can be objects of the multiplicative operator. Meanwhile an uncertainty appears because of a rest index. To overcome the uncertainty one need to consider all the 3 projections of an octonion equation.

## 1. Introduction

It is known that quaternions have one complex splitting way only i.e. a real quaternion is multiplied by a complex number. But octonions have four ways: one complex and three quaternion ways.

Here we shall consider just one quaternion way.

At the beginning we take 8-dimensional vector space over the field of real numbers, here the algebra of base vector units isomorphic to the algebra of real octonion.

$\mathbf{e}_0, \mathbf{e}_1, \mathbf{e}_2, \mathbf{e}_3, \mathbf{e}_4, \mathbf{e}_5, \mathbf{e}_6, \mathbf{e}_7,$

$\mathbf{e}_\alpha \mathbf{e}_\beta = \mathbf{e}_\gamma \varepsilon_{\alpha\beta\gamma} - \mathbf{e}_0 \delta_{\alpha\beta}$, here $\varepsilon_{\alpha\beta\gamma} = 1$, when $\alpha\beta\gamma = 123, 145, 246, 347, 176, 572, 536$;  (1)

$\delta_{\alpha\beta}$ is the Kronecker symbol.

**Definition 1.1** Let $i_1, i_2, i_3$ be the generators of the algebra of quaternions. If we multiplied these generators by unit vectors of the 8-dimensional vector space (1) we thereby should obtain new **split** vector space.

$\mathbf{u}_0 = \mathbf{e}_0, \mathbf{u}_1 = i_1 \mathbf{e}_1, \mathbf{u}_2 = i_1 \mathbf{e}_2, \mathbf{u}_3 = i_1 \mathbf{e}_3, \mathbf{u}_4 = i_2 \mathbf{e}_4, \mathbf{u}_5 = i_3 \mathbf{e}_5, \mathbf{u}_6 = i_2 \mathbf{e}_6, \mathbf{u}_7 = i_3 \mathbf{e}_7,$  (2)

here $i_1 \neq \mathbf{e}_1, i_2 \neq \mathbf{e}_2, i_3 \neq \mathbf{e}_3$.

These vectors' algebra isomorphic to **split octonion** algebra which satisfies to following multiplication table I, where $i = i_1$:

**Table I**

---

$\mathbf{u}_\alpha^2 = \mathbf{u}_0, \alpha = 0,\ldots,7, \mathbf{u}_0 \mathbf{u}_\beta = \mathbf{u}_\beta, \beta = 1,\ldots, 7,$

$\underline{\mathbf{u}_1 \mathbf{u}_2 = i\mathbf{u}_3}, \underline{\mathbf{u}_1 \mathbf{u}_3 = -i\mathbf{u}_2}, \mathbf{u}_1 \mathbf{u}_4 = \mathbf{u}_5, \mathbf{u}_1 \mathbf{u}_5 = \mathbf{u}_4, \mathbf{u}_1 \mathbf{u}_6 = -\mathbf{u}_7, \mathbf{u}_1 \mathbf{u}_7 = -\mathbf{u}_6,$

$\underline{\mathbf{u}_2 \mathbf{u}_3 = i\mathbf{u}_1}, \mathbf{u}_2 \mathbf{u}_4 = i\mathbf{u}_6, \mathbf{u}_2 \mathbf{u}_5 = i\mathbf{u}_7, \mathbf{u}_2 \mathbf{u}_6 = -i\mathbf{u}_4, \mathbf{u}_2 \mathbf{u}_7 = i\mathbf{u}_5,$

$\mathbf{u}_3 \mathbf{u}_4 = \mathbf{u}_7, \mathbf{u}_3 \mathbf{u}_5 = \mathbf{u}_6, \mathbf{u}_3 \mathbf{u}_6 = \mathbf{u}_5, \mathbf{u}_3 \mathbf{u}_7 = \mathbf{u}_4,$

$\mathbf{u}_4 \mathbf{u}_5 = \mathbf{u}_1, \underline{\mathbf{u}_4 \mathbf{u}_6 = i\mathbf{u}_2}, \mathbf{u}_4 \mathbf{u}_7 = \mathbf{u}_3,$

$\mathbf{u}_5 \mathbf{u}_6 = \mathbf{u}_3, \underline{\mathbf{u}_5 \mathbf{u}_7 = i\mathbf{u}_2},$

$\mathbf{u}_6 \mathbf{u}_7 = -\mathbf{u}_1.$

---

*Here anticommutative pairs are marked with underline, other pairs are commutative.*

**Proposition 1.2** You can now admit that namely ternaries which have element $\mathbf{u}_2$ have imaginary unit. Moreover the marked anticommutative pairs belong to ternaries which have $\mathbf{u}_2$. Thus, we can say that $\mathbf{u}_2$ is the **peculiar** element. Also from the table we can note that such expressions as $\mathbf{u}_\alpha \mathbf{u}_\beta = \mathbf{u}_\gamma$, $\mathbf{u}_\alpha \mathbf{u}_\gamma = \mathbf{u}_\beta$ for commutative pairs conserve their sign, without of these expressions: $\mathbf{u}_2 \mathbf{u}_4 = i\mathbf{u}_6$, $\mathbf{u}_2 \mathbf{u}_6 = -i\mathbf{u}_4$, therefore we can say that $\mathbf{u}_6$ is the **semi-peculiar** element. (It means that peculiarity of $\mathbf{u}_6$ is not so 'strong' as $\mathbf{u}_2$ has. Just it 'helps' for $\mathbf{u}_2$ to change the sign)

**Definition 1.3** We shall determine the norm N(u) of the split octonion by $N(u) = \sqrt{RE(uu^*)}$, where u* is conjugate octonion, its elements of indexes from 1 to 7 have inverse sign; and RE (uu*) is real and scalar part of the product.

Having this determination identity $|ab|^2 = |a|^2 |b|^2$ also satisfies the case of split quaternions, but not at our case.

*It is known that this identity can't satisfy the split quaternions without of definition 1.3*

## 2. Cubic matrices

Here author thinks that if an object of 4-dimensional space is a projection of an object of hyperspace, then an equation for 4-dimensional space is a projection of an equation for hyperspace, hence classic Pauli's matrices are projections of cubic their analog. Selecting suitable objects he got the following volumetric matrices.

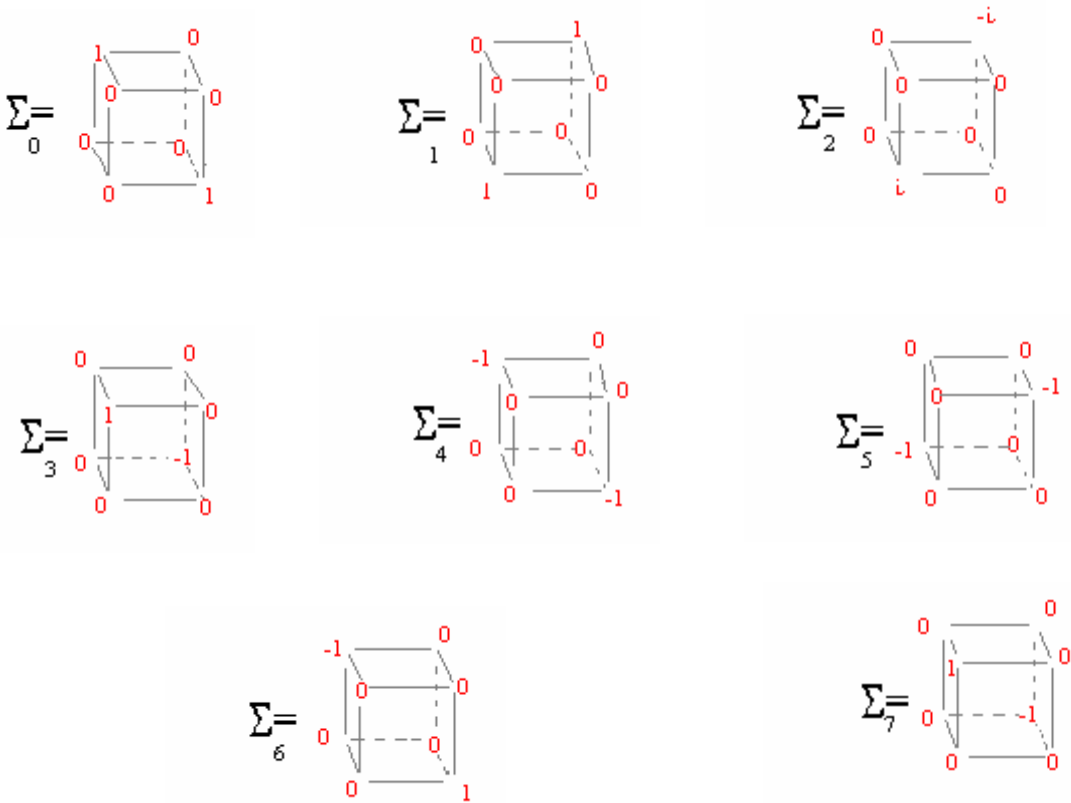

The square projections of these matrices we shall obtain with an imaginary beam parallel the following axes. (Here a number of an axis correspond to the number of the projection)

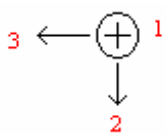 — for every matrices, except peculiar.

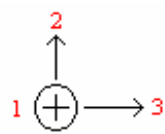 —for $\Sigma_2$

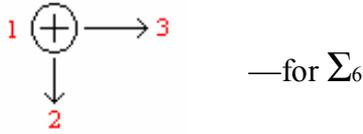 —for $\Sigma_6$

Here symbol of $\otimes$ means direction from ourselves. A projection of a matrix we shall mark with Roman numeral at right and above. **An example**:

$$\Sigma_0^{I} = \sigma_0 = \begin{pmatrix} 1 & 0 \\ 0 & 1 \end{pmatrix}, \quad \Sigma_1^{I} = \sigma_1 = \begin{pmatrix} 0 & 1 \\ 1 & 0 \end{pmatrix}, \quad \Sigma_2^{I} = \sigma_2 = \begin{pmatrix} 0 & -i \\ i & 0 \end{pmatrix}, \quad \Sigma_3^{I} = \sigma_3 = \begin{pmatrix} 1 & 0 \\ 0 & -1 \end{pmatrix}$$

As you see the first (the main) projections of matrices from $\Sigma_0$ to $\Sigma_3$ give us classical (ordinary) sigma matrices.

### 3. Properties of the cubic matrices

**Definition (!)** Now we introduce the rule: changing multiplier positions we turn the matrices for 180 degrees round one of the 3 axes. Every turned matrix we mark with tilde. The number of axe of turning we mark with Arabic figure at left and above. **An example**:

$$^{1}\tilde{\Sigma}_3^{II} = \begin{pmatrix} -1 & 0 \\ 0 & 1 \end{pmatrix}, \text{ here } \Sigma_3^{II} = \begin{pmatrix} 0 & -1 \\ 1 & 0 \end{pmatrix}$$

It means that $\Sigma_3$ was turned round the 1$^{st}$ its axis.

So, we can **distribute** projections of matrices among vector ternaries:
1) $u_1^{I}$, $u_2^{I}$, $u_3^{I}$; 2) $u_1^{II}$, $u_4^{I}$, $u_5^{I}$; 3) $u_2^{II}$, $u_4^{II}$, $u_6^{I}$; 4) $u_3^{II}$, $u_4^{III}$, $u_7^{I}$;
5) $u_1^{III}$, $u_7^{II}$, $u_6^{II}$; 6) $u_5^{II}$, $u_7^{III}$, $u_2^{III}$; 7) $u_5^{III}$, $u_3^{III}$, $u_6^{III}$. (3)

**Remark 3.1** Here in (3) in the 1$^{st}$ ternary, for instance, we get only first projections. In the 2$^{nd}$ ternary you see we get the 2$^{nd}$ projection of the 1$^{st}$ element and 1$^{st}$ projections of other elements i.e. $\Sigma_1^{II}$, $\Sigma_4^{I}$ and $\Sigma_5^{I}$. In the 4$^{th}$ ternary 3$^{rd}$ element has 2$^{nd}$ projection, 4$^{th}$ element has 3$^{rd}$ projection and 7$^{th}$ element has 1$^{st}$ projection i.e. $\Sigma_3^{II}$, $\Sigma_4^{III}$ and $\Sigma_7^{I}$.
Here I took cubic matrices as elements to be more obvious.

Therefore we can construct multiplication Table II of the cubic sigma matrices:

### Table II

1) $\Sigma_1^{I}\Sigma_2^{I} = i\Sigma_3^{I}$, $\quad ^{2}\tilde{\Sigma}_2^{I}\,^{2}\tilde{\Sigma}_1^{I} = -i\Sigma_3^{I}$, $\quad \Sigma_2^{I}\Sigma_3^{I} = i\Sigma_1^{I}$,

$^{2}\tilde{\Sigma}_3^{I}\,^{3}\tilde{\Sigma}_2^{I} = -i\Sigma_1^{I}$, $\quad \Sigma_3^{I}\Sigma_1^{I} = i\Sigma_2^{I}$, $\quad ^{3}\tilde{\Sigma}_1^{I}\,^{3}\tilde{\Sigma}_3^{I} = -i\Sigma_2^{I}$;

2) $\Sigma_1^{II} \Sigma_4^{I} = \Sigma_5^{I}$, $^3\tilde{\Sigma}_4^{I} \,^3\tilde{\Sigma}_1^{II} = \Sigma_5^{I}$, $\Sigma_4^{I} \Sigma_5^{I} = \Sigma_1^{II}$,

$^3\tilde{\Sigma}_5^{I} \,^3\tilde{\Sigma}_4^{I} = \Sigma_1^{II}$, $\Sigma_1^{II} \Sigma_5^{I} = \Sigma_4^{I}$, $^3\tilde{\Sigma}_5^{I} \,^3\tilde{\Sigma}_1^{II} = \Sigma_4^{I}$;

3) $\Sigma_2^{II} \Sigma_4^{II} = i\Sigma_6^{I}$, $^3\tilde{\Sigma}_4^{II} \,^3\tilde{\Sigma}_2^{II} = i\Sigma_6^{I}$, $^1\tilde{\Sigma}_4^{II} \,^2\tilde{\Sigma}_6^{I} = i\Sigma_2^{II}$,

$\Sigma_6^{I} \Sigma_4^{II} = -i\Sigma_2^{II}$, $\underline{^1\tilde{\Sigma}_2^{II} \,^2\tilde{\Sigma}_6^{I} = -i\Sigma_4^{II}}$, $\underline{^3\tilde{\Sigma}_6^{I} \,^3\tilde{\Sigma}_2^{II} = -i\Sigma_4^{II}}$;

4) $\Sigma_3^{II} \Sigma_4^{III} = \Sigma_7^{I}$, $^1\tilde{\Sigma}_4^{III} \,^1\tilde{\Sigma}_3^{II} = \Sigma_7^{I}$, $^2\tilde{\Sigma}_4^{III} \,^2\tilde{\Sigma}_7^{I} = \Sigma_3^{II}$,

$\Sigma_7^{I} \Sigma_4^{III} = \Sigma_3^{II}$, $\Sigma_7^{I} \Sigma_3^{II} = \Sigma_4^{III}$, $^1\tilde{\Sigma}_3^{II} \,^2\tilde{\Sigma}_7^{I} = \Sigma_4^{III}$;

5) $\Sigma_1^{III} \Sigma_6^{II} = -\Sigma_7^{II}$, $^3\tilde{\Sigma}_6^{II} \,^2\tilde{\Sigma}_1^{III} = -\Sigma_7^{II}$, $^3\tilde{\Sigma}_6^{II} \,^3\tilde{\Sigma}_7^{II} = -\Sigma_1^{III}$,

$\Sigma_7^{II} \Sigma_6^{II} = -\Sigma_1^{III}$, $\Sigma_1^{III} \Sigma_7^{II} = -\Sigma_6^{II}$, $^3\tilde{\Sigma}_7^{II} \,^2\tilde{\Sigma}_1^{III} = -\Sigma_6^{II}$;

6) $\Sigma_5^{II} \Sigma_2^{III} = i\Sigma_7^{III}$, $^2\tilde{\Sigma}_2^{III} \,^3\tilde{\Sigma}_5^{II} = i\Sigma_7^{III}$, $\Sigma_2^{III} \Sigma_7^{III} = i\Sigma_5^{II}$,

$^2\tilde{\Sigma}_7^{III} \,^2\tilde{\Sigma}_2^{III} = i\Sigma_5^{II}$, $^3\tilde{\Sigma}_5^{II} \,^3\tilde{\Sigma}_7^{III} = i\Sigma_2^{III}$, $\Sigma_7^{III} \Sigma_5^{II} = -i\Sigma_2^{III}$;

7) $\Sigma_5^{III} \Sigma_6^{III} = \Sigma_3^{III}$, $^2\tilde{\Sigma}_6^{III} \,^2\tilde{\Sigma}_5^{III} = \Sigma_3^{III}$, $\Sigma_6^{III} \Sigma_3^{III} = \Sigma_5^{III}$,

$^1\tilde{\Sigma}_3^{III} \,^2\tilde{\Sigma}_6^{III} = \Sigma_5^{III}$, $\Sigma_5^{III} \Sigma_3^{III} = \Sigma_6^{III}$, $^2\tilde{\Sigma}_3^{III} \,^2\tilde{\Sigma}_5^{III} = \Sigma_6^{III}$

---

**Remark 3.2** Let us observe for instance point 5) of the table II. Here you see multiplication rules for matrices $\Sigma_1$, $\Sigma_6$ and $\Sigma_7$. According to distribution (3) we take into account following projections: $\Sigma_1^{III}$, $\Sigma_6^{II}$ and $\Sigma_7^{II}$. Let us need to multiply $\Sigma_1$ by $\Sigma_6$, we see from the line 1 of the table I that $u_1 u_6 = -u_7$ i.e. $\Sigma_1^{III} \Sigma_6^{II} = -\Sigma_7^{II}$

To get $\Sigma_6 \Sigma_1$, it was empirical obtained that according to the rule (!) we may turn cube of $\Sigma_6$ round its 3rd axis and we may turn cube of $\Sigma_1$ round its 2nd axis, so we get $^3\tilde{\Sigma}_6^{II} \,^2\tilde{\Sigma}_1^{III} = -\Sigma_7^{II}$ etc.

Having a projection of matrix the product we restore the whole cubic matrix of the product.
Also we can see that product of two marked with underline peculiar matrices is peculiar too, i.e. if we multiply $\Sigma_2$ by $\Sigma_6$ or vice versus, we must turn these matrices.

**Remark 3.3** One understands it was described properties of multiplicative operator $\odot$, such as $\Sigma_i \odot \Sigma_j = u_{ijk} \Sigma_k$, here $u_{ijk}= 1$ when ijk = 145, 154, 347, 356, 365, 374, 451, 473, 563,

and $u_{ijk}= -1$ when ijk = 167, 176, 671; also:

$u_{ijk}= i$ when ijk = 123, 231, 246, 257, 275, 312, 462, 572

and $u_{ijk}= -i$ when ijk = 246

*Proof of nonassociativity*

Let us prove the property of nonassociativity. It needs to take 3 elements which don't belong to the same ternary.

Let us for instance need to multiply $\Sigma_1$ by $\Sigma_2$ and $\Sigma_6$. We see that $\Sigma_1$ and $\Sigma_2$ are elements of 1st ternary, but $\Sigma_6$ doesn't belong to it. From another hand $\Sigma_2$ and $\Sigma_6$ are elements of 3rd ternary, but $\Sigma_1$ doesn't belong to it. Thus $\Sigma_1$, $\Sigma_2$ and $\Sigma_6$ don't belong to the same ternary, which satisfies our case.

a) We get product of $(\Sigma_1 \Sigma_2)\Sigma_6$. First step: we multiply $\Sigma_1$ by $\Sigma_2$, from the point 1) of the table II we have $\Sigma_1^I \Sigma_2^I = i\Sigma_3^I$, i.e. we get $i\Sigma_3$. Second step: we multiply $\Sigma_3$ by $\Sigma_6$, from the point 7) of the table II we have $^1\tilde{\Sigma}_3^{III}\,^2\tilde{\Sigma}_6^{III} = \Sigma_5^{III}$, i.e. we get $\Sigma_5$. Taking into account '$i$' we obtain $(\Sigma_1 \Sigma_2)\Sigma_6 = i\Sigma_5$.

b) We get product of $\Sigma_1(\Sigma_2 \Sigma_6)$. First step: we multiply $\Sigma_2$ by $\Sigma_6$, from the point 3) of the table II we have $^1\tilde{\Sigma}_2^{II}\,^2\tilde{\Sigma}_6^{I} = -i\Sigma_4^{II}$, i.e. we get $-i\Sigma_4$. Second step: we multiply $\Sigma_1$ by $\Sigma_4$, from the point 2) of the table II we have $\Sigma_1^{II} \Sigma_4^{I} = \Sigma_5^{I}$, i.e. we get $\Sigma_5$. Taking into account '-$i$' we obtain $\Sigma_1(\Sigma_2 \Sigma_6) = -i\Sigma_5$.

But the case of **a)** has another result. □

**Remark 3.4** Nevertheless it is necessary to pay attention to the fact that only projections of the octonion matrices are subjects to the multiplicative operation. The cubic matrices cannot be multiplied because they have 3 indices. Therefore the square projection only can be multiplied. Meanwhile because of a rest index we have an uncertainty, to eliminate it we have to consider all the 3 projections of an octonion equation.

**Remark 3.5** One can see that the given space is not isotropic. Here are peculiar directions. Only matrix $\Sigma_2$ that corresponds to the peculiar element #2, has '$i$' in it, other matrices have just real units. But the peculiarity of $\Sigma_2$ can be observed among other 6 matrices. For the 4-dimensional space the second element is not peculiar.

## 4. Supposed gamma-matrices form

Here we'll try to construct octonion gamma-matrices

**The first case**. In [1] one can see: $\Gamma_1 = -\sigma_1 \otimes \sigma_1 \otimes \sigma_2$, here the symbol of $\otimes$ means direct product. If extrapolate it, we try to construct then following matrices:

$\Gamma_1 = -\Sigma_1 \otimes \Sigma_1 \otimes \Sigma_2$, $\Gamma_2 = -\Sigma_1 \otimes \Sigma_2 \otimes \Sigma_2$, $\Gamma_3 = -\Sigma_1 \otimes \Sigma_3 \otimes \Sigma_2$,

$\Gamma_4 = -\Sigma_2 \otimes \Sigma_4 \otimes \Sigma_2$, $\Gamma_5 = -\Sigma_3 \otimes \Sigma_5 \otimes \Sigma_2$, $\Gamma_6 = -\Sigma_2 \otimes \Sigma_6 \otimes \Sigma_2$,
$\Gamma_7 = -\Sigma_3 \otimes \Sigma_7 \otimes \Sigma_2$ and $\Gamma_0 = 1$ (4)

**The second case.** In [2] one can see: $\gamma^0 = \sigma_1 \otimes \sigma_0$, hence we have: $\Gamma_0 = \Sigma_0 \otimes \Sigma_1 \otimes \Sigma_0$
Also $\gamma^l = i\sigma_2 \otimes \sigma_l$, here $l = 1, 2, 3$; hence we have: $\Gamma_l = i\Sigma_1 \otimes \Sigma_2 \otimes \Sigma_l$, and
$\Gamma_4 = i\Sigma_2 \otimes \Sigma_2 \otimes \Sigma_4$, $\Gamma_5 = i\Sigma_3 \otimes \Sigma_2 \otimes \Sigma_5$, $\Gamma_6 = i\Sigma_2 \otimes \Sigma_2 \otimes \Sigma_6$, $\Gamma_7 = i\Sigma_3 \otimes \Sigma_2 \otimes \Sigma_7$, (5)

## 5. General form of a volumetric Dirac equation solution

Finally, looking back at the hypothesis of volumetric matrices, author tries to get general form of a volumetric equation solution.
Landau and Lifshitz (see [2]) and other say that the solution of the Dirac equation is:

$$\Psi = const * \exp(-i/\hbar(Et - \bm{p}\bm{r})) \qquad (6)$$

Let us make a little of transformation:
let $(E, \bm{p})$ be 4-dimensional momentum and $(t, \bm{r})$ be 4-dimensional interval.
And let $P = (E + u_1 p_1 + u_2 p_2 + u_3 p_3)$, $R = (t + u_1 r_1 + u_2 r_2 + u_3 r_3)$ and
$D = RE(PR^*)$ (see **Definitions 1.1 and 1.3** here)
Consequently one can write (6) as:
$$\Psi = \exp(a_0 + (-iD/\hbar)) = \exp(a_0 + i(-D/\hbar)) \qquad (7)$$
*here and after $a_i$ is a real number.*

Let now be $-D/\hbar = S$, thus one can write (6) as:

$$\Psi = \exp(a_0 + iS) \qquad (8)$$

And finally, considering volumetric Dirac equation (see (4) and (5) here) one can realize that its solution can be written as:

$$\Psi = \exp(a_0 + ia_1 S_1 + ja_2 S_2 + ka_3 S_3) \qquad (9)$$

*Here $i, j, k$ are generators of quaternion algebra, $S_1$ is the solution of the 1st projection of the volumetric Dirac equation, $S_2$ is the solution of the 2nd projection of this equation and $S_3$ is the solution of the 3rd projection of this equation.*

Therefore one can say that the volumetric equation solution is a **quaternion construction**.

## 6. Conclusion

As square matrices cannot be nonassociative the author introduced cubic sigma-matrices of the octonion Dirac equation and multiplicative operator for the projections. Also he described gamma-matrices hypothetical form and general form of volumetric Dirac equation solution.
From the present work it follows that the hyperspace is not isotropic. It has 24 dimensions and it is divided into 3 parts, each part has 8 dimensions.

It is reasonable to ask whether these 3 hyperspace parts connect to the quark colors in the quantum chromodynamics?

The results may be interested for hyperspace research and hyperspace features of particles.